# Experimental Study of the Shortest Reset Word of Random Automata

Evgeny Skvortsov†, Evgeny Tipikin‡
skvortsoves@gmail.com, etipikin@gmail.com

†Google Inc., USA; ‡ Ural State University, Yekaterinburg, Russia

**Abstract.** In this paper we describe an approach to finding the shortest reset word of a finite synchronizing automaton by using a SAT solver. We use this approach to perform an experimental study of the length of the shortest reset word of a finite synchronizing automaton. The largest automata we considered had 100 states. The results of the experiments allow us to formulate a hypothesis that the length of the shortest reset word of a random finite automaton with $n$ states and 2 input letters with high probability is sublinear with respect to $n$ and can be estimated as $1.95n^{0.55}$.

## 1 Introduction

A *deterministic finite automaton* (DFA) is a triple $\mathcal{A} = (Q, \Sigma, \delta)$, where $Q$ is a set of states, $\Sigma$ is an input alphabet, and $\delta : Q \times \Sigma \to Q$ is a transition function defining an action of the letters in $\Sigma$ on $Q$. We use a common concise notation denoting $\delta(\dots\delta(\delta(\mathbf{q}, a_0), a_1), \dots a_k)$ by $\mathbf{q}a_0 \dots a_k$.

A word $w \in \Sigma^*$ is said to be a *reset word* for a DFA $\mathcal{A}$ if its action leaves $\mathcal{A}$ in one particular state no matter what state it starts at: $\mathbf{q}_1 w = \mathbf{q}_2 w$ for all $\mathbf{q}_1, \mathbf{q}_2 \in Q$. A DFA $\mathcal{A}$ is called *synchronizing* if it possesses a reset word. In this paper we describe results of an experimental study of the length of the shortest reset word of random automata.

It can be easily shown that if an automaton with $n$ states is synchronizing then it has a reset word of length less than $n^3$. However, the tightness of this bound is far from obvious. In 1964, Černý formulated a conjecture concerning the upper bound of the length of the shortest reset word of a synchronizing DFA [5]: the length cannot be larger than $(n-1)^2$. By now the Černý conjecture is arguably the longest standing open problem in the combinatorial theory of finite automata. The tightest upper bound that has been obtained so far is $(n^3 - n)/6$; it was proved by Pin [14] in 1983.

Though no bound better than cubic has been proven for the shortest reset word, most naturally occurring automata have reset words of subquadratic length. Automata with reset word of length $\Theta(n^2)$ are considered to be exceptional. For a long time the only infinite series of such automata was the one proposed by Černý [5]. The other substantially different ones [1,2] have only recently been constructed.

There are several theoretical and experimental results that support the statement that most synchronizing automata have a relatively short reset word. First, Higgins [10] has shown that the composition of $2n$ random mappings of a set of size $n$ into itself *with high probability* (whp) is a mapping with an image of size 1. (By "high probability" we mean that the probability tends to 1 as $n$ goes to infinity.) In terms of automata, Higgins's result means that a random automaton with an alphabet of size larger than $2n$ whp has a reset word of length $2n$. Indeed, if we pick an automaton uniformly at random among all automata with $n$ states and $2n$ letters, then the action of a word composed of all the letters is identical to a mapping composed of $2n$ random mappings. Later it was shown [16] that a random automaton with $n$ states over an alphabet of size $n^{0.5+\varepsilon}$ has a reset word of quadratic length with high probability for any $\varepsilon > 0$.

The probability distribution of the length of the shortest reset word of a random automaton can be studied experimentally for small $n$. It is unlikely that there is a polynomial algorithm that can find the shortest reset word in general case because the problem belongs to $FP^{NP[log]}$ [13], which means that the problem is both NP-hard and co-NP-hard. Moreover, approximating the length of the shortest reset word has also been shown to be hard [3]. Nevertheless, it is possible that the problem restricted to a certain class of automata (for instance see [**?**]) or to random automata is easy and can be successfully solved by an appropriate heuristic. Recently, Roman [15] has developed a genetic algorithm for finding a short reset word and in particular, applied it to random automata. In this paper we present the results of applying of SAT solvers to the problem of finding the shortest reset word.

SAT (or Boolean Satisfiability) is a combinatorial problem of finding a boolean assignment that satisfies a given boolean formula in conjunctive normal form. SAT was one of the first problems proven to be NP-complete [6]. The development of practical algorithms for solving instances of SAT (so called SAT-solvers) is an area of active research and there is a regular competition of these algorithms. These days the problems that participate in SAT competitions have hundreds of thousands of variables and millions of literals. This is especially surprising when one recalls that SAT is NP-complete. This observation does not formally contradict the NP-hardness of SAT, but shows that hard instances of SAT rarely occur in practice. There are various approaches to explaining this phenomenon in greater detail [11,7,4].

SAT is also known to be a natural language for a variety of combinatorial problems. In this paper we show that the problem of finding the shortest reset word of a finite automaton can be naturally reduced to a few SAT instances. We apply a SAT solver to those instances and recover the reset word from the resulting boolean assignment.

As mentioned, Roman [15] was using a genetic algorithm to find a reset word of random automata. Since genetic algorithms are incomplete, the results of [15] allow one to assume only an upper bound on the length of the shortest reset word. It turns out that even for an alphabet of size 2 as the number of states grows, the probability of the automata being synchronizing approaches 1. In this paper we also study automata over a 2-letter alphabet. It is easy to see that if the size of the alphabet gets larger, the length of the shortest reset word of a random automata decreases.

We were able to find the shortest reset words of randomly generated automata with up to 100 states. We argue that the results of our experiments are a reasonable basis for the hypothesis of the length of the shortest reset word of a random automaton. The hypothesis is given in the following formula:

$$\ell(n) \approx 1.95n^{0.55},$$

where $n$ is the number of states of the random automaton and $\ell(n)$ is the length of the shortest reset word.

The rest of the paper is organized as follows. In Section 2, we describe how the problem of finding the shortest reset word can be reduced to a collection of instances of SAT. In Section 3, we formally define the notion of a random automaton. In Section 4, we present results of experiments and what we believe they mean. We conclude in Section 5 with a short discussion.

## 2 Solving Automata Synchronization Problem via Reduction to SAT

Given a finite automaton $\mathcal{A} = (Q, \{a, b\}, \delta)$ and an integer $c$, we build a 3-CNF formula $\phi_A^c$ such that $\phi_A^c$ is satisfiable if and only if $\mathcal{A}$ has a reset word $w$ of length $c$. We denote the prefix of $w$ of length $t$ by $w|_{1\ldots t}$.

The formula $\phi_A^c$ contains two types of variables:

– For each $t \in 1, \ldots, c$, we introduce a variable $u_t$. Setting $u_t$ to *true* is interpreted as "the $t$-th letter of $w$ is $a$" and setting $u_t$ to *false* is interpreted as "the $t$-th letter of $w$ is $b$".
– For each $\mathbf{q} \in Q$ and $t \in \{0, \ldots, c\}$, we introduce a variable $x_{\mathbf{q}t}$. A variable $x_{\mathbf{q}0}$ is used to mark whether an automaton can be initially in a state $\mathbf{q}$ or not. When $t \neq 0$, setting $x_{\mathbf{q}t}$ to *false* is interpreted as "there does not exist a state $\mathbf{u}$ such that $\mathbf{u}w|_{1\ldots t} = \mathbf{q}$". It is convenient for us to interpret setting $x_{\mathbf{q}t}$ to *true* as "there *may* exist a state $\mathbf{u}$ such that $\mathbf{u}w|_{1\ldots t} = \mathbf{q}$". In other words, we will enforce setting $x_{\mathbf{q}t}$ to *true* and will not enforce *false*.

There are $c$ variables of the first type and $(c+1)n$ variables of the second type. Therefore the resulting boolean formula contains $(c+1)n + c$ boolean variables.

There are also three types of clauses in $\phi_A^c$:

– For each $\mathbf{q} \in Q$ we assert that initially the automaton can be in this state by adding a one literal clause
$$x_{\mathbf{q}0}.$$
– For each $\mathbf{q} \in Q$ and $t \in \{0, \ldots, c-1\}$ we add the following elementary disjunctions to $\phi_A^c$:
$$\neg x_{\mathbf{q}t} \vee \neg u_t \vee x_{(\mathbf{q}a)(t+1)},$$

$$x_{\mathbf{q}t} \vee u_t \vee x_{(\mathbf{q}b)(t+1)}.$$

Note that these disjunctions are equivalent to the following implications:

$$x_{\mathbf{q}t} \wedge u_t \to x_{(\mathbf{q}a)(t+1)}, \tag{1}$$

$$x_{\mathbf{q}t} \wedge \neg u_t \to x_{(\mathbf{q}b)(t+1)}. \tag{2}$$

The clauses of the first and the second types together enforce setting $x_{\mathbf{q}t}$ to *true* if and only if the state $\mathbf{q}$ can be achieved from some state of $\mathcal{A}$ by applying the prefix $w|_{1\ldots t}$.

– For each 2-element subset $\{\mathbf{p}, \mathbf{q}\} \subseteq Q$, where $\mathbf{p} \neq \mathbf{q}$ we add the following elementary disjunctions to $\phi_A^c$:

$$\neg x_{\mathbf{q}c} \vee \neg x_{\mathbf{p}c}. \tag{3}$$

The clauses of the third type ensure that at most one of the variables $x_{\mathbf{q}c}$ may be true.

If $w$ is a reset word of length $c$ for $\mathcal{A}$, then the formula $\phi_A^c$ is satisfiable. Indeed the satisfying assignment is obtained as follows. Values of the variables $u_1, \ldots, u_c$ are determined by reading the word $w$ and setting $u_t$ to *true* or *false* according to the value of the $t$-th letter of $w$. Then we assign $x_{\mathbf{q}0} = true$ for all $\mathbf{q} \in Q$. Next, for each $t = 1, \ldots, k$ and for each $\mathbf{q} \in Q$, we assign $x_{\mathbf{q}t}$ to *true* if it must be done to satisfy some clause of type (1) or (2). Otherwise, we assign $x_{\mathbf{q}t}$ to $false$. It is easy to see that after such an assignment for any $t$ and $\mathbf{q}$ we have $x_{\mathbf{q}t}$ equal to true if and only if $\mathbf{q} = \mathbf{u}w|_{1\ldots t}$, for some $\mathbf{u} \in Q$. Since $w$ is a synchronizing word, all clauses of type (3) are satisfied. Analogously, if the formula $\phi_A^c$ is satisfiable, then the values of the variables $u_1, \ldots, u_c$ in the satisfying assignment define a word $w$ of length $c$, and the fact that all clauses of $\phi_A^c$ are satisfied implies that $w$ is a reset word.

There are $n$ clauses of the first type, $2cn$ clauses of the second type and $\frac{n(n-1)}{2}$ clauses of the third type. In total we have $\frac{n(n-1)}{2} + n(2c+1)$ clauses. Clauses of the first type have one literal, clauses of the second type have three literals and clauses of the third type have two literals each. Therefore the formula $\phi_A^c$ contains $n^2 + 6cn$ literals in total.

Thus, we can use a SAT solver to answer the question,

"Can $\mathcal{A}$ be synchronized by a word of length $c$?"

We use MiniSAT solver [9] to find the solution to this problem. SAT algorithms development is a very active research area and each year new solvers win the competition. MiniSAT was developed in 2003 and has become a state-of-the-art algorithm since then. The algorithm is relatively simple and yet very efficient — its performance is comparable to the best present day solvers. In some years, the SAT competition has a specialized MiniSAT-hack tournament. For more details on the algorithm see [8,9].

Once we have an algorithm that can check whether there is a reset word of given length we can find the length of the shortest reset word by performing binary search (see Fig. 1). Note that there is a polynomial algorithm for checking whether $\mathcal{A}$ is synchronizing [5]. Thus, we use the algorithm in Fig. 1 only for synchronizing automata.

```
INPUT: Synchronizing n state automata A, we assume that n > 0
OUTPUT: The shortest reset word of A
r = n^3
l = 0
while True:
  # Note: we use integer division in the next formula.
  c = (l+r)/2
  if c == l:
    return SynchroWord(r)
  if SynchroWord(c) is not None:
    r = c
  else:
    l = c
```

**Fig. 1.** Binary search applied to the shortest reset word problem. We assume that we have a function SynchroWord($c$) that returns a reset word of length less than or equal to $c$ if such word exists and returns $None$ otherwise.

## 3 Random Automaton

In the experimental section we study the length of the shortest reset word of a random automaton over a 2-letter alphabet. Formally, Random Automaton $\mathbb{A}(n)$ with $n$ states over an alphabet $\Sigma$ can be defined as a discrete probability space $(\Omega_{\mathbb{A}}, P)$, where sample space $\Omega_{\mathbb{A}}$ is the set of all automata over $\Sigma$. To define a specific automaton $\mathcal{A} = (Q, \Sigma, \delta)$ one needs to define $\delta(\mathbf{q}, a)$, for each $\mathbf{q} \in Q$ and $a \in \Sigma$. Thus, it is easy to see that $|\Omega_{\mathbb{A}}| = n^{|\Sigma|n}$. We set the probability of all elements of the sample space to be equal, and consequently for all $\mathcal{A}$ we have $P(\mathcal{A}) = n^{-|\Sigma|n}$. We also consider a probabilistic space Random Synchronizing Automaton $\mathbb{A}'(n)$. Formally, $\mathbb{A}'(n)$ is defined as a probabilistic space induced by $\mathbb{A}(n)$ on the event "$\mathcal{A}$ is synchronizing".

The length of the shortest reset word $\ell(n)$ is a random variable over the probabilistic space $\mathbb{A}'(n)$. To study the behaviour of the random variable $\ell(n)$ as $n$ tends to infinity we define the expectation of $\ell(n)$ by $r(n)$ and the variance of $\ell(n)$ by $d(n)$, that is

$$r(n) = \mathbf{E}\left(\ell(n)\right),$$

$$d(n) = \mathbf{V}\left(\ell(n)\right).$$

Note that while $\ell(n)$ is a random variable for each $n$, the functions $r(n)$ and $d(n)$ are deterministic.

## 4 Experimental Results

We performed a series of experiments for different $n$, where $n$ is the number of states in the automaton. For a given $n$, the experiment consists of the following.

We generate a random automaton with $n$ states. Then we check whether this automaton is synchronizing and if so, we find a reset word for this automaton using the algorithm described in Fig. 1. Then we record the result of synchronization, i.e., whether the automaton is synchronizing and the length of the shortest reset word.

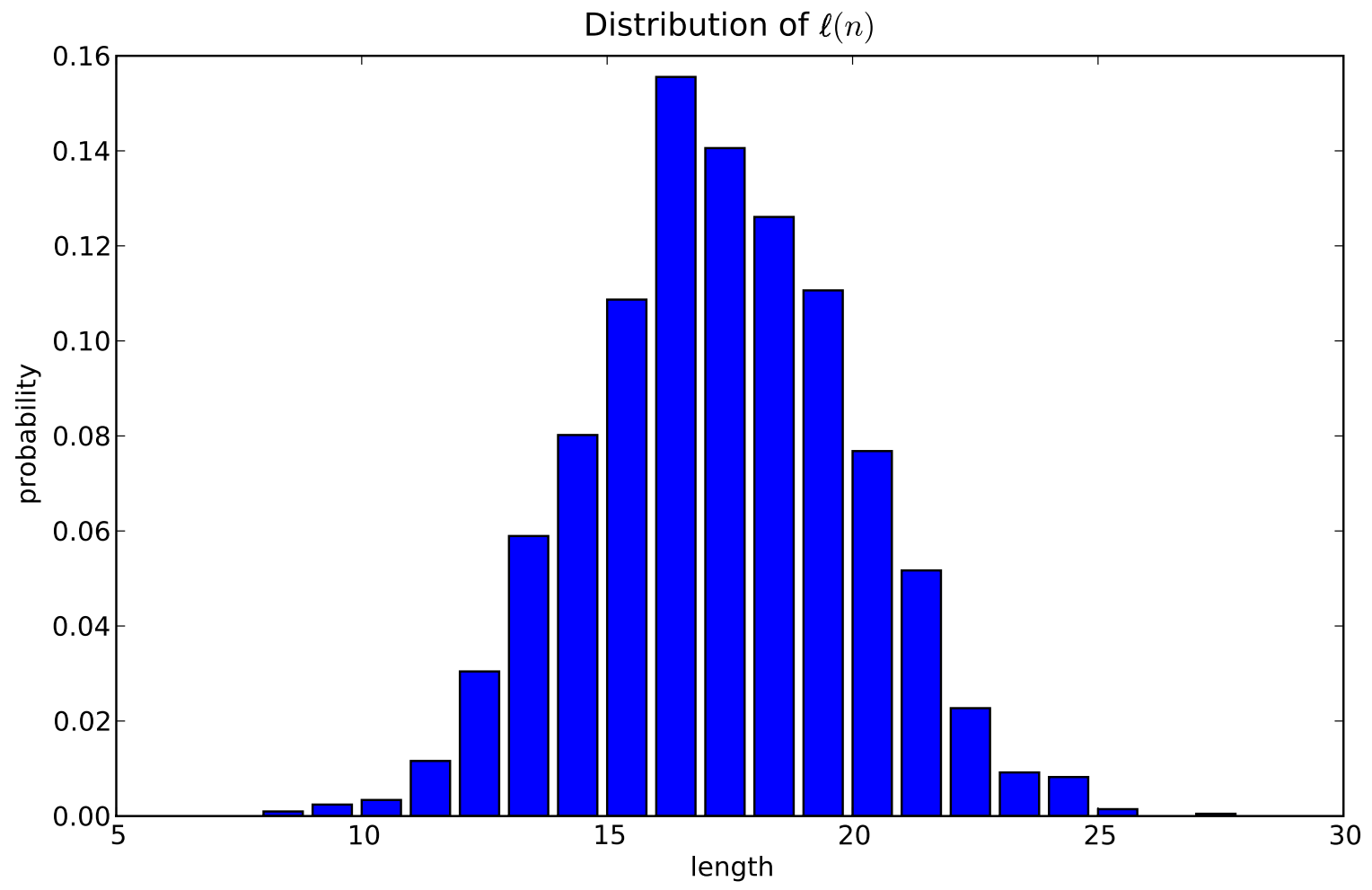

**Fig. 2.** Probability distribution of $\ell(50)$.

For a specified number of states $n$, we performed a number of such experiments. The larger $n$ is the more time it takes to solve the problem of finding the reset word, so for larger $n$ we performed fewer experiments. For each $n \in \{1, 2, \ldots, 20, 25, 30, \ldots, 50\}$ we performed 2000 experiments, for each $n \in \{55, 60, 65, 70\}$ we performed 500 experiments and for $n \in \{75, 80, \ldots, 100\}$ we performed 200 experiments. In our experiments we used a personal computer with an Intel(R) Core(TM)2 Duo P8600 2.4GHz CPU and 4GB of RAM. The program for calculations was written in Java. The average calculation time was 2.7 seconds for $n = 50$ and 70 seconds for $n = 100$.

Thus, for each value of $n$ participated in experiments we have an approximated probabilistic distribution of $\ell(n)$ and an estimated probability of the event "$\mathbb{A}(n)$ is synchronizing". In Fig. 2 we show the distribution of $\ell(50)$.

### 4.1 Synchronization of $\mathbb{A}$

The larger $n$ is, the larger the fraction of generated random automata that are synchronizing. For $n = 100$ only 1 out of 200 automata that we generated happened to not be synchronizing. Thus, we conclude that it is likely that

$$P(\text{“}\mathbb{A}\text{ is synchronizing”}) \underset{n\to\infty}{\longrightarrow} 1.$$

### 4.2 Expectation of $\ell(n)$

It appears that the function $r(n)$ follows a certain trend. To check whether the dependence of the mean value of the distribution $\ell(n)$ follows a power law, we plot the graph in log/log space in Fig. 3. From the graph we conclude that it is a combination of some effects that are present for small $n$ and an affine function that is obeyed for large $n$. To extract the behaviour of $\mathbb{A}$ for large $n$, we ignore data points for $n < 20$. We use the least squares method to find an affine function that best reflects the dependency of $\log(r)$ on $\log(n)$. We find that

$$\log(r(n)) \approx 0.55 \log(n) + 0.67. \tag{4}$$

Taking the exponent of both sides of (4) we obtain the equation

$$r(n) \approx 1.95n^{0.55} \tag{5}$$

In Fig. 4, we plot the graph of $r$ versus $n$ and the curve given by (5). It is interesting to note that the obtained approximation starts to fit the data at $n = 17$, approximating some data points that were not used in training.

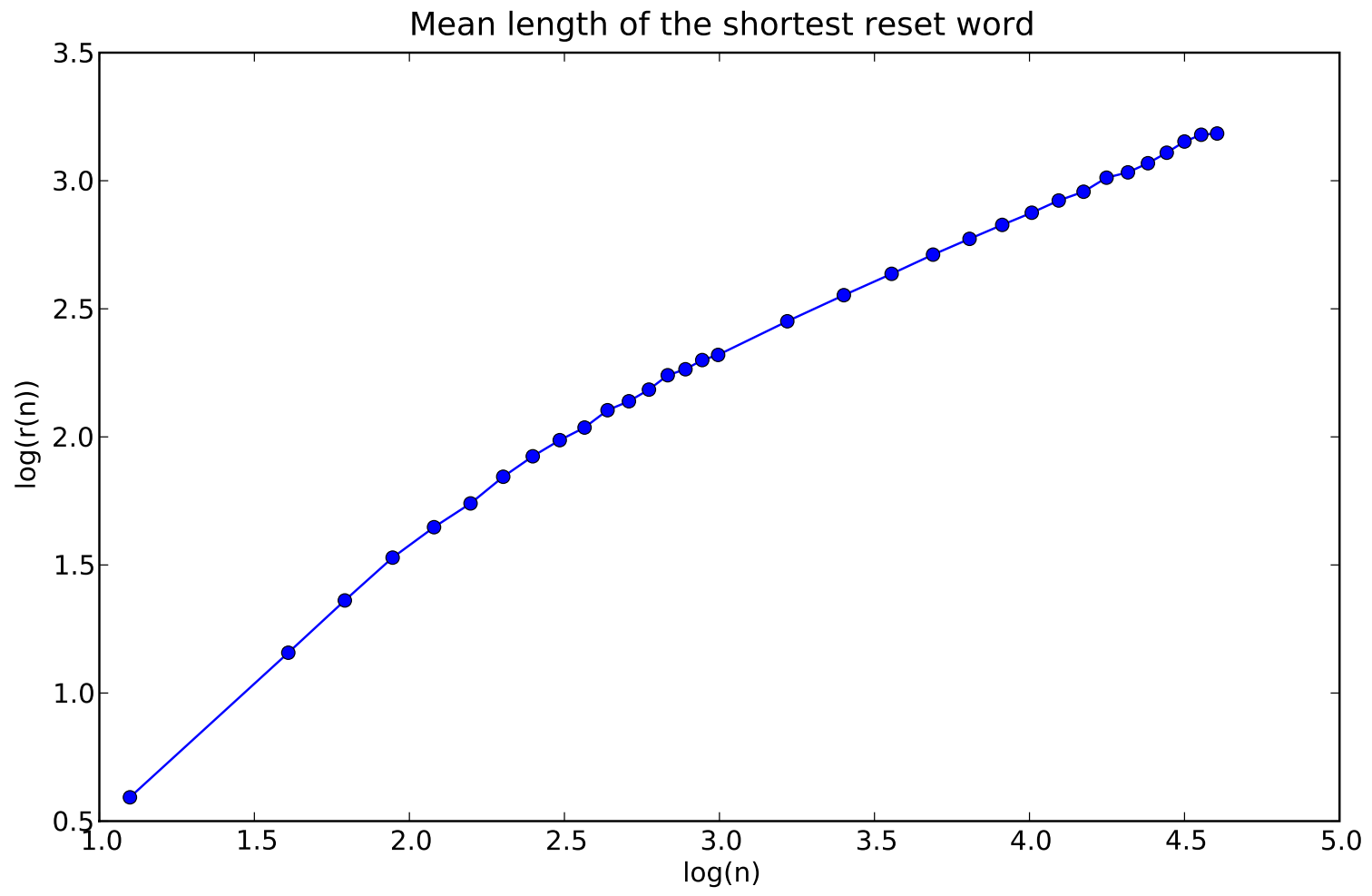

**Fig. 3.** The graph of the logarithm of the number of states of automata $n$ versus the logarithm of the length of the shortest reset word $r$.

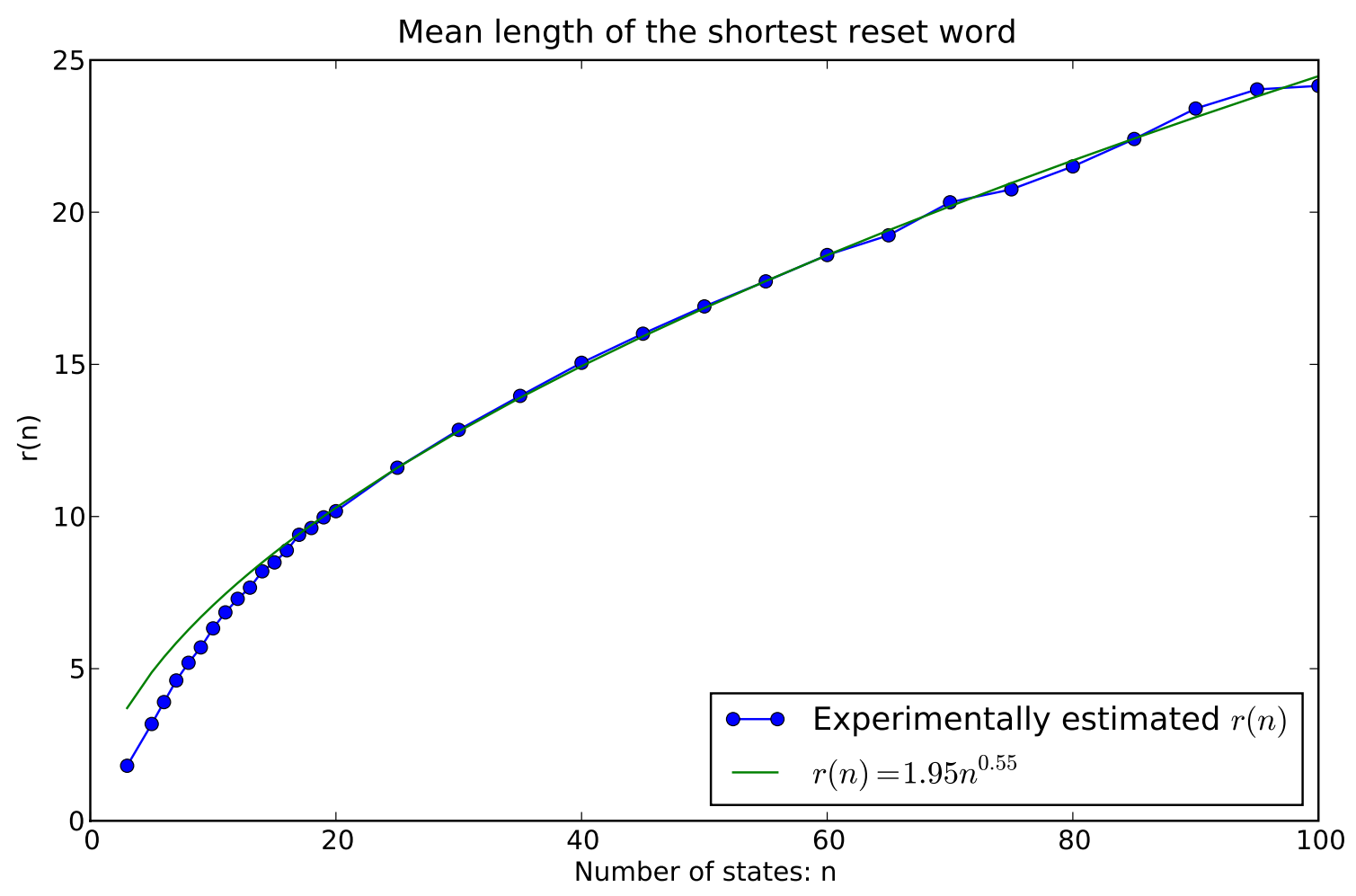

**Fig. 4.** The graph of the mean length of the shortest reset word versus the number of states of the random automata and a power function approximating it.

### 4.3 Variance of $\ell(n)$

Recall that we denote variance of $\ell(n)$ by $d(n)$. Our experiments show that as $n$ grows, $d(n)$ also grows. But what is more interesting to look at is the behaviour of the function $\frac{\sqrt{d(n)}}{r(n)}$. In Fig. 5, we show the graph of $\frac{\sqrt{d(n)}}{r(n)}$, which appears to tend to 0 as $n$ goes to infinity. Below we discuss what that means for the distribution of $\ell(n)$.

The Chebyshev inequality reads (we omit the parameter $n$ for conciseness):

$$\forall M > 0 \; P(|\ell - r| \geq M\sqrt{d}) \leq \frac{1}{M},$$

and after transformation we have

$$\forall M > 0 \; P\left(\left|\frac{\ell}{r} - 1\right| \geq M\frac{\sqrt{d}}{r}\right) \leq \frac{1}{M}. \tag{6}$$

We have $\frac{\sqrt{d(n)}}{r(n)} \underset{n\to\infty}{\longrightarrow} 0$ so there exists $M(n) \underset{n\to\infty}{\longrightarrow} \infty$ such that

$$M(n)\frac{\sqrt{d(n)}}{r(n)} \underset{n\to\infty}{\longrightarrow} 0.$$

Therefore, we have

$$P(\ell(n) = r(n) + o(r(n))) \underset{n\to\infty}{\longrightarrow} 1.$$

In other words, with high probability $\ell(n)$ is approximately equal to $r(n)$.

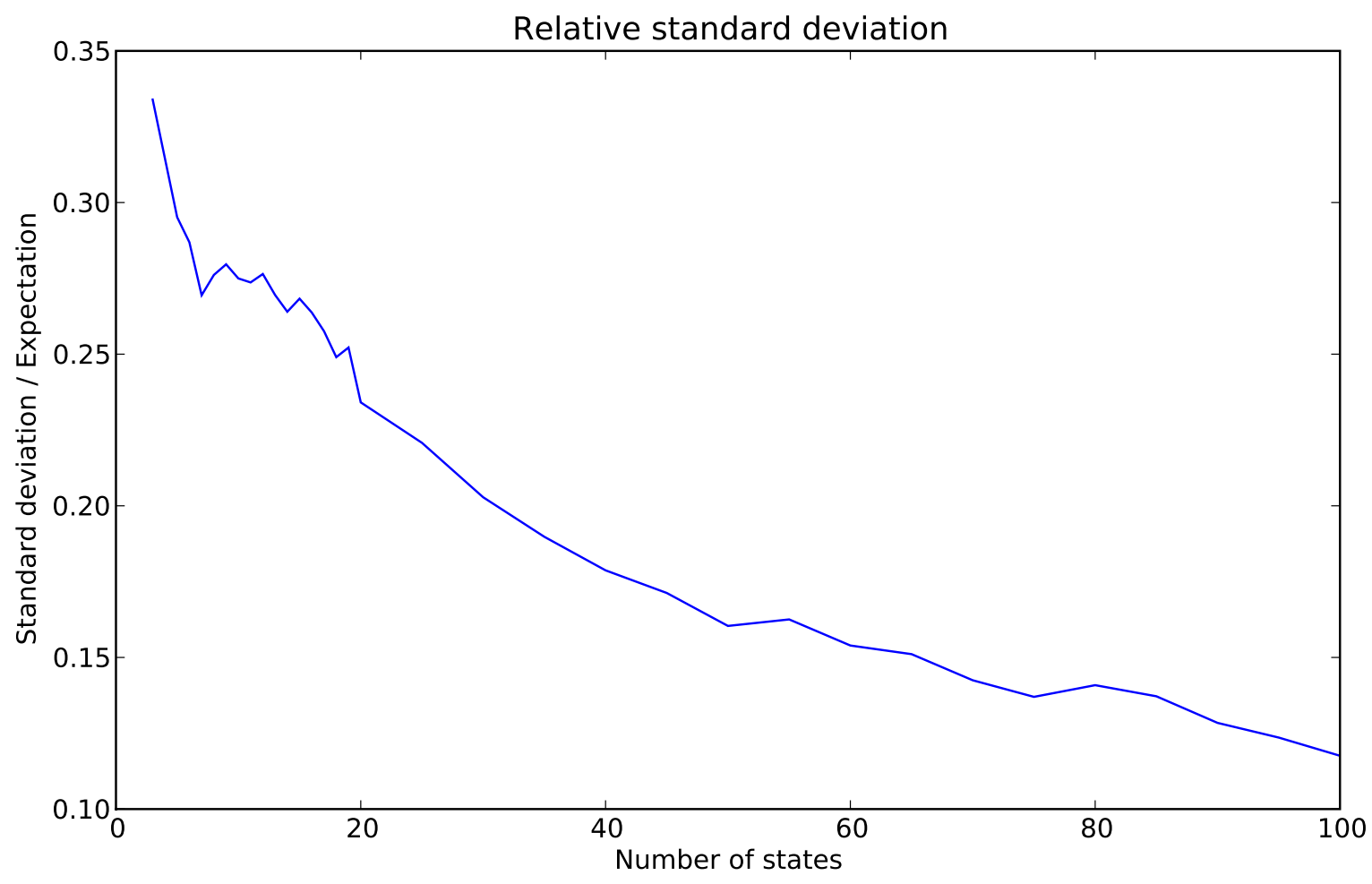

**Fig. 5.** The graph of $\frac{\sqrt{d(n)}}{r(n)}$ appears to tend to 0 as $n$ goes to infinity.

## 5 Conclusion and Discussion

We interpret experimental results as indicating that as $n$ goes to infinity, a random automaton is synchronizing with high probability. Also with high probability the the length of its shortest reset word can be computed as

$$\ell(n) \approx 1.95n^{0.55}. \tag{7}$$

In particular, we believe that the experimental data we obtained suggests that the length of the shortest synchronizing word of a random automaton is sublinear with respect to the number of states.

It worth noting that our conclusion (7) directly contradicts a conjecture that Roman formulated in [15]. Namely, Roman conjectured that the mean length of the shortest reset word for a random $n$-state synchronizing automaton is almost equal to $0.486n$. Roman's experiments with random automata consisted of two parts: for each $n = 5, 6, \ldots, 14$ one thousand random $n$-state automata were generated and then for each $n = 15, 16, \ldots, 100$ ten random $n$-state automata were generated. The linear estimate $\ell(n) \approx 0.486x + 1.654$ was suggested on the basis of the results of the first part of the experiments and then it was extrapolated even though the reported results of the second part did not really support the extrapolation. In contrast, we believe that both our and Roman's experiments with larger $n$ indicate that a random automaton is synchronized by a word of length sublinear with respect to the number of states.

We are also aware of another series of experiments with random automata synchronization performed by Gusev (these experiments are mentioned in [1]).

A direct comparison of our results with those by Gusev is impossible because he used a different random automata model. However, on a qualitative level our conclusions tend to quite agree with Gusev's.

**Acknowledgement.** We are grateful to Prof. M.V. Volkov for numerous productive discussions on the topic, and to the anonymous reviewers for their remarks which have helped us make the article more accurate and clear.